\documentclass{iau}
\usepackage{graphicx}

\title[Searching for IMBHs in GGCs] 
{Searching for IMBHs in Galactic globular clusters through radial velocities of individual stars}

\author[B. Lanzoni]   
{Barbara Lanzoni
}

\affiliation{Department of Physics and Astronomy, University of
  Bologna, \,Viale Berti Pichat 6/2, 40127- Bologna, Italy \,email:
  {\tt barbara.lanzoni3@unibo.it} \\[\affilskip] }

\pubyear{2015}
\volume{xxx}  
\pagerange{...-....}
\jname{Star Clusters and Black Holes in Galaxies across cosmic time}
\editors{A.C. Editor, B.D. Editor \& C.E. Editor, eds.}
\begin{document}

\maketitle

\begin{abstract}
I present an overview of our ongoing project aimed at building a new
generation of velocity dispersion profiles ad rotation curves for a
representative sample of Galactic globular clusters, from the the
radial velocity of hundreds individual stars distributed at different
distances from the cluster center. The innermost portion of the
profiles will be used to constrain the possibile presence of
intermediate-mass black holes. The adopted methodology consists in
combining spectroscopic observations acquired with three different
instruments at the ESO-VLT: the adaptive-optics assisted, integral
field unit (IFU) spectrograph SINFONI for the innermost and highly
crowded cluster cores, the multi-IFU spectrograph KMOS for the
intermediate regions, and the multi-fiber instrument
FLAMES/GIRAFFE-MEDUSA for the outskirts.  The case of NGC 6388,
representing the pilot project that motivated the entire program, is
described in some details.

\end{abstract}

\firstsection 
\section{Introduction}
Confirming the existence of intermediate mass ($10^3-10^4 M_\odot$)
black holes (IMBHs) would have a dramatic impact on a number of open
astrophysical problems, ranging from the formation of supermassive BHs
and their co-evolution with galaxies, to the origin of ultraluminous
X-ray sources in nearby galaxies, up to the detection of gravitational
waves (e.g. \cite[Gebhardt et al. 2005]{gebh05}).  However, the
evidence gathered so far in support of the existence of IMBHs are
inconclusive and controversial (see, e.g. \cite[Noyola et
  al. 2010]{noyola10}, \cite[Anderson \& van der Marel
  2010]{jay10}). Globular clusters (GCs) are thought to be the best
places where to search for these elusive objects. In fact, the
extrapolation of the "Magorrian relation" (\cite[Magorrian et
  al. 1998]{magorrian}) down to the IMBH masses naturally leads to the
GC mass regime. Moreover, numerical simulations have shown that the
cores of dense star clusters are the ideal habitat for the formation
of IMBHs (e.g. \cite[Portegies Zwart et al. 2004]{portegies04}).  For
these reasons, the recent years have seen an increasing number of
works dedicated to the search for IMBHs in GCs, exploiting all
observational channels, ranging from the detection of X-ray and radio
emission (see \cite[Strader et al. 2012]{strader12} and \cite[Kirsten
  \& Vlemmings 2012]{kirsten12}, and references therein), to the
detailed study of the shape of GC density and velocity dispersion
profiles (e.g., \cite[Gebhardt et al. 2000]{gebh00}; \cite[Gerssen et
  al. 2002]{gerssen02}, \cite[Lanzoni et al. 2007]{lanz07imbh},
\cite[Noyola et al. 2010]{noyola10}, \cite[L{\"u}tzgendorf et
  al. 2012]{lutz12}).

In spite of such an effort, however, no firm conclusions could be
drawn to date.  This is mainly because of the great difficulties
encountered from both the theoretical and the observational points of
view.

\section{Methodology}
\label{sec:method}
To search for IMBHs in Galactic GCs (GGCs) we follow the ``dynamical
approach'', i.e., we perform detailed studies of the cluster structure
and dynamics with the goal of identifying the central density and
velocity dispersion (VD) cusps theoretically predicted for stellar
systems hosting an IMBH (e.g. \cite[Baumgardt, Makino, \& Hut
  2005]{baumg05}; \cite[Miocchi 2007]{miocchi07}).

Despite its importance, very little is empirically known to date about
GGC internal dynamics, especially in the most crowded central regions.
Recent results suggest that some insight can be obtained from the
observations of exotic stellar populations, like blue straggler stars
and millisecond pulsars (e.g.,\cite[Ferraro et al
  2003]{fer03},\cite[Ferraro et al. 2009a]{fer09m30}, \cite[Ferraro et
  al. 2009b]{fer09ter}, \cite[Ferraro et al. 2012]{dyn_clock}).
However, a detailed knowledge of the VD profile and rotation curve is
still missing in the vast majority of GGCs because of observational
difficulties.  The velocity components on the plane of the sky can be
obtained from internal proper motion measurements. However, this
requires high-precision photometry and astrometry on quite long
temporal baselines, together with precise estimates of GC distances.
While this is starting to be feasible (thanks to the combination of
multi-epoch HST observations and the improved techniques of data
analysis; \cite[Anderson \& van der Marel 2010]{jay10}; \cite[Massari
  et al. 2013]{massari13}, \cite[Bellini et al 2014]{bellini14}), it
is still very challenging in the high density central regions, where
the crucial dynamical information constraining the presence of an IMBH
is expected.  The line of sight (\emph{los}) velocity is in principle
easier to obtain (through spectroscopy) and measurable in any cluster
region and in GCs at any distance within the Galaxy. In practice,
however, the standard approach commonly used in extra-Galactic
astronomy (i.e. measuring the line broadening of integrated light
spectra) is prone to a severe ``shot noise bias'': if a few bright
stars are present, the acquired spectrum is completely dominated by
their light and the line broadening is therefore not produced by the
true VD of the underlying stellar population (e.g. \cite[Dubath,
  Meylan \& Mayor 1997]{du97}).  The alternative approach is to
measure the dispersion about the mean of the radial velocities of
statistically significant samples of individual stars. While this is
safe from obvious biases, it is observationally challenging,
especially in the highly crowded centers of GCs, requiring
multi-object and spatially resolved spectroscopy.

Motivated by the astrophysical importance of determining the
kinematical properties of GGCs, and because of the known shot noise
bias affecting the integrated light spectroscopy method, we designed
an ambitious project aimed at measuring the radial velocity of
resolved stars along the entire radial extension of a representative
sample of GGCs, by means of a multi-instrument approach:
adaptive-optics (AO) assisted spectroscopy with SINFONI in the
innermost, highly crowded regions, KMOS and FLAMES observations in the
intermediate and external regions, respectively.  Two Large Programmes
are currently running at the Very Large Telescope of the European
Southern Observatory (Prop. ID: 193.D-0232 and 195.D-0750, PI:
F. R. Ferraro), collecting data for a representative sample of $\sim
30$ GGCs.  Here we present the results obtained for the pilot project
on NGC 6388 (see \cite[Lanzoni et al. 2013]{lanz13} and \cite[Lapenna
  et al. 2015]{lape15}).

\begin{figure}[!t]
\begin{center} 
\includegraphics[scale=0.55]{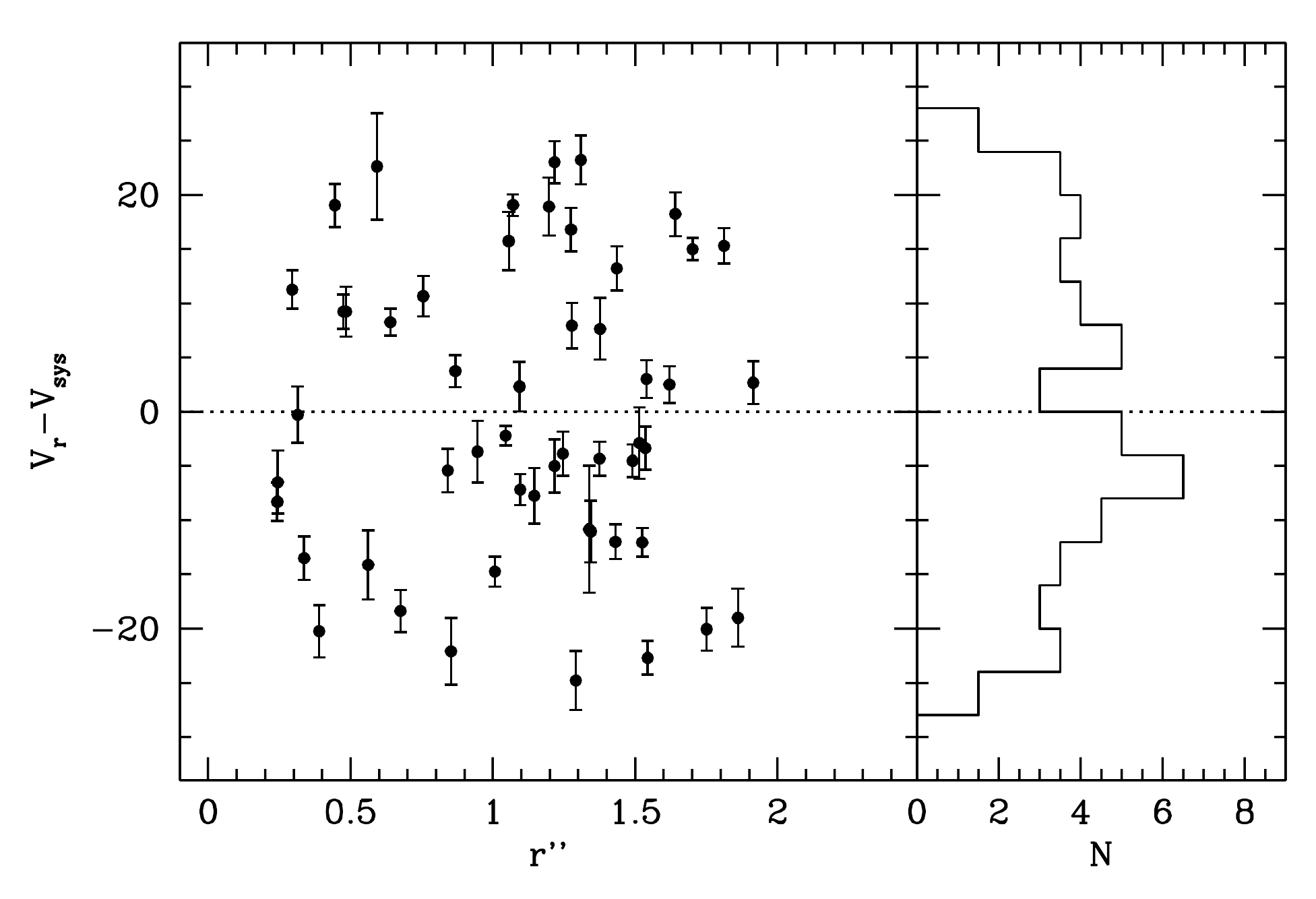}
\caption{{\it Left panel:} distribution of the radial velocities
  measured for the 52 SINFONI targets in NGC 6388, referred to the
  cluster systemic velocity $V{\rm sys}$ and plotted as function of
  the distance from the centre.  {\it Right panel:} histogram of the
  radial velocity distribution. The apparent bimodality indicates the
  presence of systemic rotation.}
\label{vrot_sinfo}
\end{center}
\end{figure}

\begin{figure}[!t]
\begin{center} 
\includegraphics[scale=0.55]{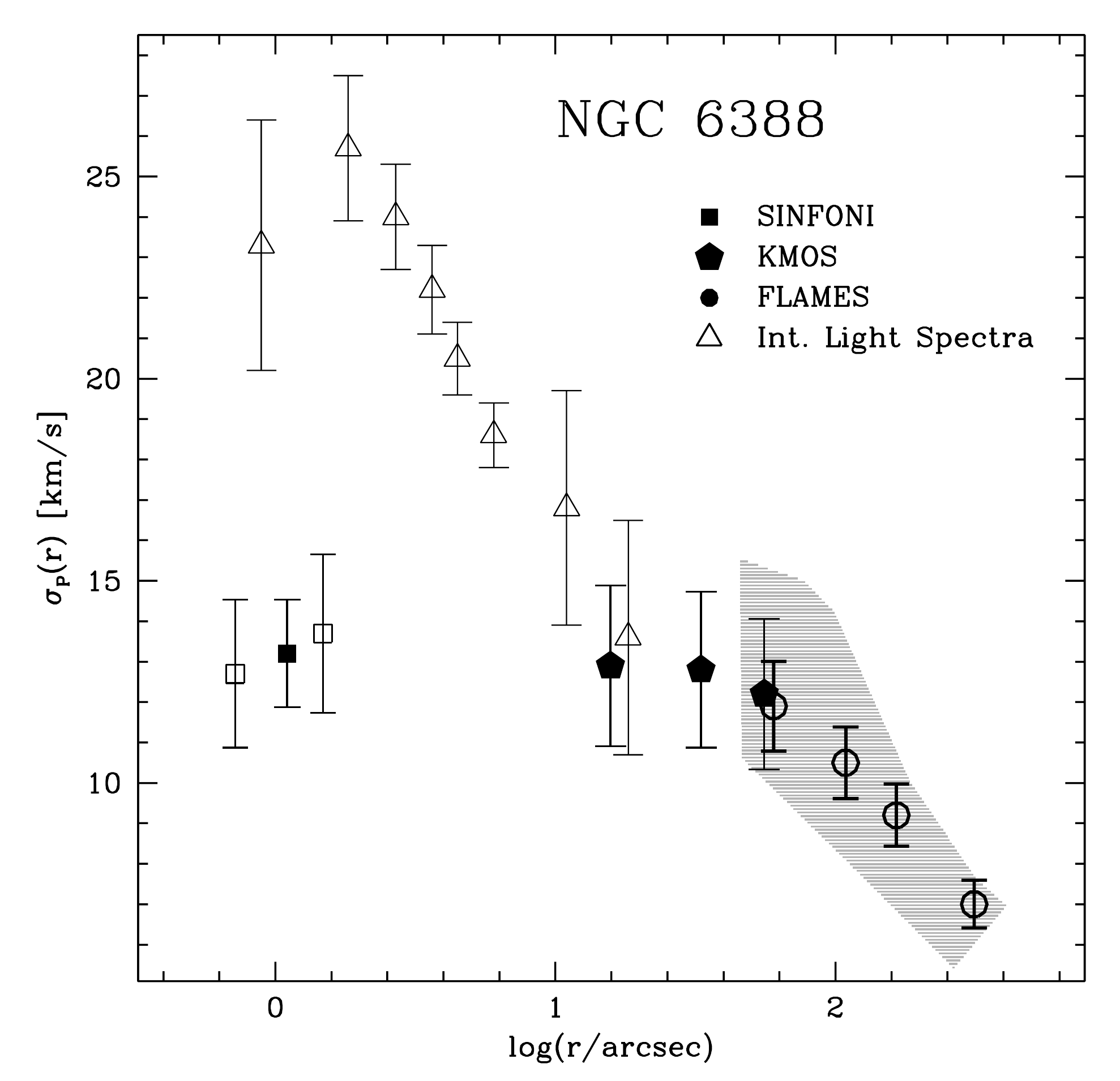}
\caption{Line-of-sight velocity dispersion profile of NGC 6388
  computed from the radial velocities of $\sim 400$ individual stars
  as measured with with SINFONI (squares), KMOS (pentagons), and
  FLAMES (empty circles). The the gray region indicates the dispersion
  obtained for different choices of the radial bins.  The velocity
  dispersion profile obtained from integrated-light spectra
  (\cite[L{\"u}tzgendorf et al. 2011]{lutz11}) is also shown for
  comparison (empty triangles).}
\label{vdisp}
\end{center}
\end{figure}

\begin{figure}[!t]
\begin{center}
\includegraphics[scale=0.55]{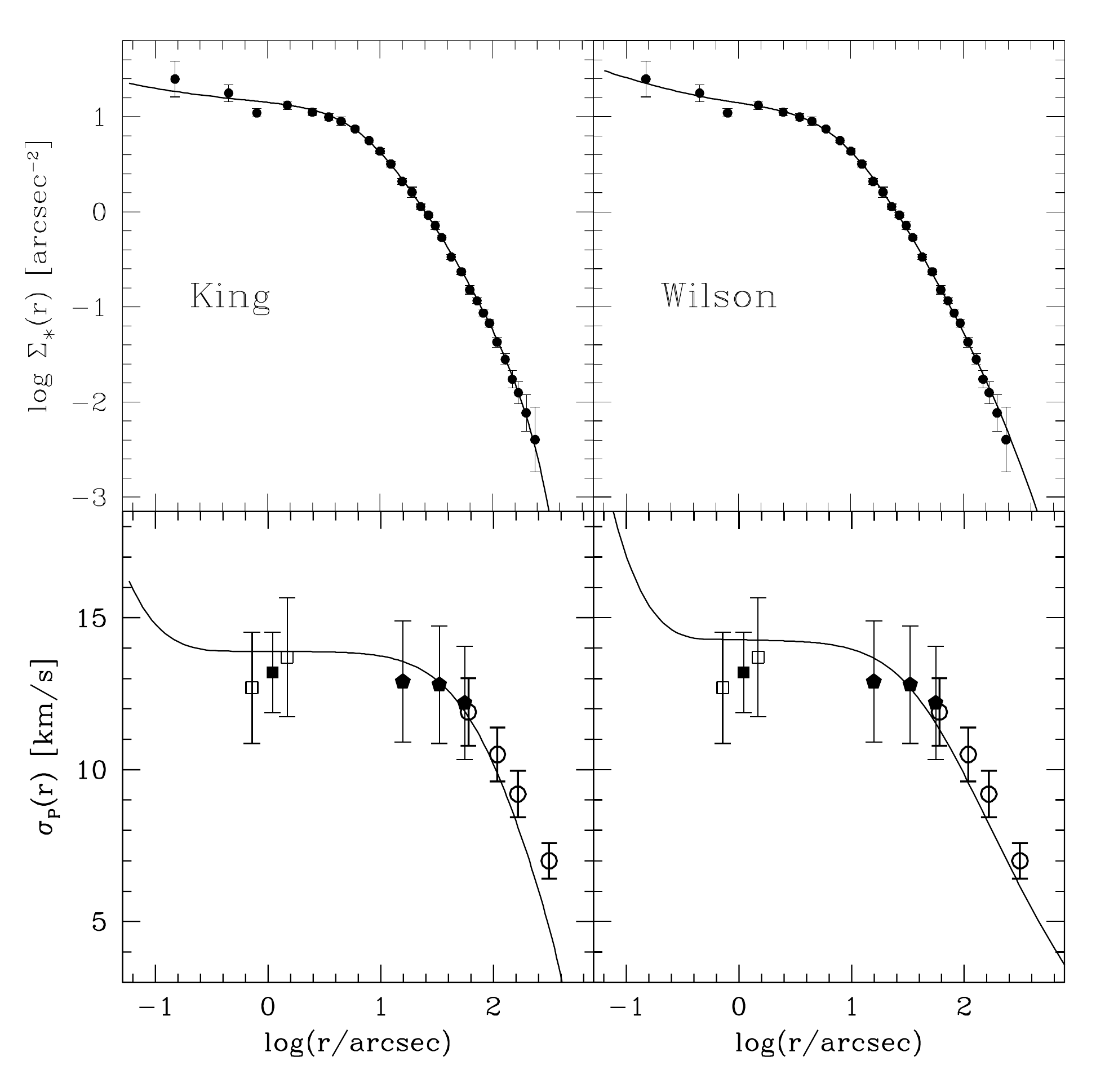}
\caption{Comparison between the observations and the best-fit
  self-consistent King and Wilson models with central IMBH (left- and
  right-hand panels, respectively). The observed density profile
  (solid circles in the upper panels) is from \cite[Lanzoni et al.
    (2007)]{lanz07imbh}. The observed velocity dispersion profiles are
  the same as in Fig. \ref{vdisp}. An IMBH mass of $\sim 2000 M_\odot$
  is assumed in these models.}
\label{profs}
\end{center}
\end{figure}
 
\begin{figure}[!t]
\begin{center} 
\includegraphics[scale=0.7]{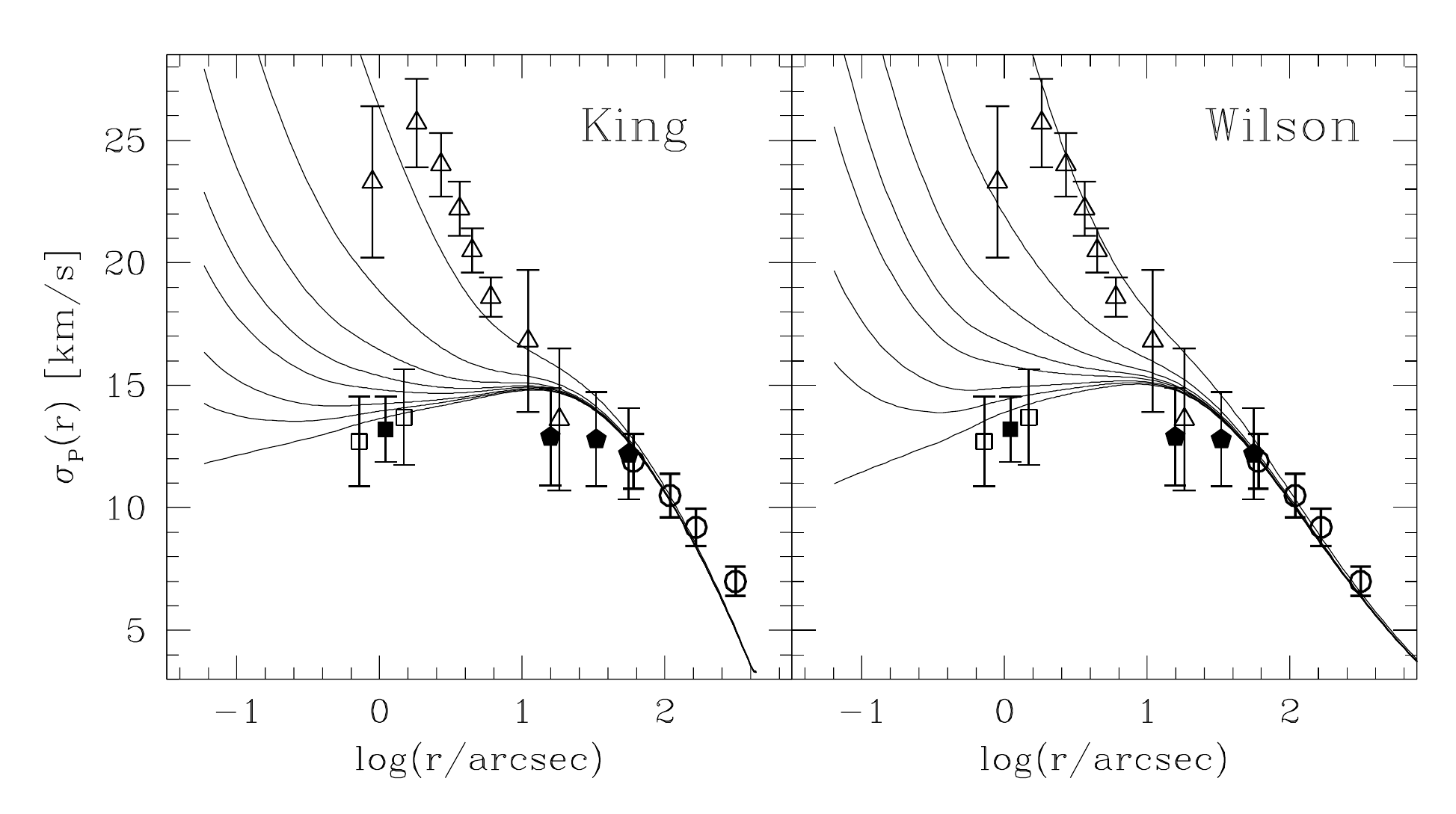}
\caption{Comparison between the observed velocity dispersion profiles
  (the same as in Fig. \ref{vdisp}) and two families of Jeans
  models. {\it Left panel:} solid lines correspond to Jeans models
  calculated from the King density profile shown in the upper-left
  panel of Fig. \ref{profs} and by assuming different black hole
  masses: from bottom to top, $M^{\rm J}_{\rm BH}/M^{\rm sc}_{\rm
    BH}=0, 0.5, 1, 2, 3, 5, 10, 30$, with $M^{\rm sc}_{\rm BH}=2147
  M_\odot$. {\it Right panel:} the same as in the left panel, but for
  Jeans models calculated from the Wilson density profile shown in the
  upper-right panel of Fig. \ref{profs}. In this case, $M^{\rm
    sc}_{\rm BH}=2125 M_\odot$}
\label{jeans} 
\end{center}
\end{figure}

\section{The pilot project: NGC 6388}
\label{sec:6388} 
By using high spatial resolution ($0.025''\times 0.027''$/pix) data
acquired with the HST/ACS High Resolution Channel (HRC), we discovered
a shallow central density cusp in the inner $1''$ of this cluster and,
from the comparison with King models including a central IMBH
(\cite[Miocchi 2007]{miocchi07}), we concluded that a $\sim 6\times
10^3 M_\odot$ BH could be hidden in NGC 6388 (\cite[Lanzoni et al.
  2007]{lanz07imbh}).
 
To further investigate this possibility we then adopted the
multi-instrument approach described above for determining the cluster
VD profile and search for the central cusp expected in the presence of
an IMBH.

~

\noindent
{\bf SINFONI -} the innermost cluster regions have been properly
investigated by exploiting the high spatial resolution capabilities of
SINFONI (\cite[Eisenhauer et al.2003]{eisen}), a near-IR (1.1-2.45
$\mu$m) integral field spectrograph fed by an AO module and mounted on
the YEPUN (VLT-UT4) telescope at the ESO Paranal Observatory.  By
using the 100 mas plate-scale and the $K$-band grating (sampling the
1.95-2.45 $\mu$m wavelength range), we acquired spectra at a
resolution $R=4000$ for $\sim 60$ stars located in a $3.2''\times
3.2''$ region centered on the cluster gravity center (as quoted in
\cite[Lanzoni et al. 2007]{lanz07imbh}).  From the $^{12}$C$^{16}$O
band-heads (\cite[Origlia et al 1997]{origlia97}) and a few atomic
lines we then measured the radial velocity ($V_r$) of 52 giant stars
located within $\sim 2''$ from the center of NGC 6388.

\noindent
{\bf KMOS - } KMOS is a second generation spectrograph equipped with
24 IFUs that can be allocated within a $7.2'$ diameter field of
view.  Each IFU covers a projected area on the sky of about
$2.8''\times2.8''$, and it is sampled by an array of 14$\times$14
spatial pixels (hereafter spaxels) with an angular size of $0.2''$
each.  KMOS is equipped with four gratings providing a maximum
spectral resolution $R$ between $\sim$ 3200 and 4200 over the 0.8-2.5
$\mu$m wavelength range.  We have used the YJ grating and observed in
the 1.00-1.35 $\mu$m spectral range at a resolution $R\approx$3400,
corresponding to a sampling of about 1.75 $\rm \mathring{A}$
pixel$^{-1}$, i.e. $\sim$ 46 km s$^{-1}$ pixel$^{-1}$ at 1.15
$\mu$m. By measuring several atomic lines for each star, we finally
attained an average uncertainty of $\sim 3$ km s$^{−1}$ in the radial
velocity estimates. With KMOS we thus measured $V_r$ for a total of 82
giant stars located within $\sim 70''$ from the center of NGC 6388.

\noindent
{\bf FLAMES - } The spectra of individual stars out to $\sim 600''$
have been acquired by using the ESO VLT multi-object spectrograph
FLAMES, in the MEDUSA UVES+GIRAFFE combined mode. This provides 132
fibers to observe an equivalent number of targets in one single
exposure, over a field of view of $25'$ in diameter.  Four pointings
of 2320 s each have been obtained with the GIRAFFE grating HR21 (which
samples the Ca II triplet spectral range at a resolution $R=16200$)
and the UVES setup Red Arm 580, covering the wavelength range $4800
\mathring{\rm A}<\lambda<6800\mathring{\rm A}$ at a resolution
$R=47000$.  The targets have been selected from the combined
HST/ACS-WFC and ESO/WFI photometric catalog discussed in \cite[Lanzoni
  et al.  (2007)]{lanz07imbh} and \cite[Dalessandro et
  al. (2008)]{ema08}, considering only isolated objects (with no
brighter stars within a circle of $1''$ radius), having $V<17$ and
being located along the canonical evolutionary sequences of the
color-magnitude diagram (CMD). By also including two additional
data-sets retrieved from the ESO Archive, we finally obtained the
radial velocities of about 270 member stars of the cluster.

In all cases, radial velocities have been measured by adopting the
Fourier cross-correlation method (\cite[Tonry et al. 1979]{tonry79})
as implemented in the {\sl fxcor} IRAF task.  The observed spectrum is
cross-correlated with a template of known radial velocity and a
cross-correlation function (i.e., the probability of correlation as a
function of the pixel shift) is computed.  Then, this is fitted by
using a Gaussian profile and its peak value is derived. Once the
spectra are wavelength calibrated, the pixel shift obtained by the
Gaussian fit is converted into radial velocity.  We employed different
reference template spectra according to the adopted spectral
configuration and stellar type.

The systemic velocity of NGC 6388 has been computed by conservatively
using all stars with radial velocities between 60 and 105 km s$^{-1}$,
deriving a value of $V_r = 82.0 \pm 0.5$ km s$^{-1}$, in agreement
with the literature (\cite[Harris 1996]{harris}).  Hereafter,
$\widetilde V_r$ will indicate radial velocities referred to the
cluster systemic velocity: $\widetilde V_r\equiv V_r-V_{\rm sys}$.

\subsection{Systemic rotation}
\label{sec:rot} 
The radial velocity distribution for the 52 SINFONI targets is plotted
in Fig. \ref{vrr_sinfo}. The histogram in the right-hand panel shows a
clear bimodality, indicating the presence of systemic rotation.  To
more quantitatively investigate this possibility, we used the method
fully described in \cite[Bellazzini et al. 2012]{bellaz12}. We
considered a line passing through the cluster center with position
angle PA varying between $0^{\rm o}$ (North direction) and $90^{\rm
  o}$ (East direction), by steps of $15^{\rm o}$.  For each value of
PA, such a line splits the SINFONI sample in two. The difference
$\Delta\langle \widetilde V_r\rangle$ between the mean radial velocity
of the two sub-samples was computed and has been found to show a
coherent sinusoidal behavior, which is a signature of rotation. The
best-fit position angle of the rotation axis is PA$_0=11^{\rm o}$ and
the amplitude of the rotation curve $A_{\rm rot}=8.5$ km s$^{-1}$.
The Kolmogorov-Smirnov probability, however, indicates a not very high
statistical significance, possibly due to the limited sample. Some
hints of rotation are found also from the KMOS and FLAMES samples, but
again the significance is weak.  Additional data symmetrically
sampling the cluster are needed before drawing any firm conclusion
about ordered rotation in NGC 6388.
 
~

\subsection{Velocity dispersion}
\label{sec:vr}
To compute the projected VD profile, the surveyed area has been
divided in a set of concentric annuli, chosen as a compromise between
a good radial sampling and a statistically significant number ($> 50$)
of stars.  In each radial bin, $\sigma_p$ has been computed from the
dispersion of the values of $\widetilde V_r$ measured for all the
stars in the annulus, by following the Maximum Likelihood method
described in \cite[Walker et al. (2006]{walker06}; see also
\cite[Martin et al. 2007]{martin07}; \cite[Sollima et
  al. 2009]{sollima09}). An iterative $3\sigma$ clipping algorithm was
applied in each bin.  The error estimate is performed by following
Pryor \& Meylan (1993).  The resulting VD profile is shown in Figure
\ref{vdisp}.  Given the number of stars in the SINFONI data set (52
objects in total) we computed $\sigma_p(r)$ by considering both one
single central bin (solid square in the figure) and two separate
annuli (26 stars at $r<1.2''$, 26 stars beyond; empty squares). For
the external data set we tried different sets of radial bins, finding
consistent results (grey area in Fig. \ref{vdisp}). The derived values
are shown in the figure as black diamonds and empty circles for the
KMOS and the FLAMES samples, respectively.  We find that the velocity
dispersion of NGC 6388 has a central ($r\sim 1''$) value of $\sim 13$
km s$^{-1}$, stays approximately flat out to $r\sim 60''$, and then
decreases to $\sim 7$ km s$^{-1}$ at $200''<r <600''$.  If no $\sigma$
clipping algorithm is applied, the VD profile remains almost
unchanged, with the only exception of the outermost data-point that
rises to $\sim 8.9\pm0.8$ km s$^{-1}$.

The inner part ($r< 10''$) of the VD profile is clearly incompatible
with the one obtained by \cite[L{\"u}tzgendorf et al. (2011]{lutz11},
hereafter L11) from the line broadening of integrated-light spectra,
which shows a steep rise toward the cluster centre, up to values of
$23-25$ km s$^{-1}$ (empty triangles in Fig. \ref{vdisp}), suggesting
the presence of an IMBH of $(1.7\pm 0.9)\times 10^4 M_\odot$.  While
the exact reason for such a disagreement is not completely clear, a
detailed comparison between our radial velocity measurements and L11
radial velocity map suggests that their measure is affected by the
shot noise bias mentioned above (see Sect. 4.1 and Fig. 12 in
\cite[Lanzoni et al. 2013]{lanz13}).

\subsection{Comparison with models}
\label{sec:models}
In order to constraint the presence and the mass of a central IMBH in
NGC 6388, we followed two different and complementary
approaches. First, starting from a family of self-consistent models
admitting a central IMBH, we selected the one that best reproduces the
observed density and VD profiles. This provided us with the
corresponding structural and kinematical parameters in physical units,
including the IMBH mass. Second, we used the observed density profile
and included a variable central point mass, to solve the spherical
Jeans equation (this is what is done also in L11). We therefore
obtained the corresponding family of VD profiles, which were then
compared to the observations to constrain the BH mass. The first
approach is more realistic in terms of the stellar mass-to-light ratio
($M/L$), since it takes into account various populations of stars with
different masses and different radial distributions (as it is indeed
expected and observed in mass segregated GCs).  Moreover it has the
advantage that the models stem from a distribution function which is
known a priori. Hence, the model consistency (i.e., the non-negativity
of the distribution function in the phase-space) is under control and
guaranteed by construction. However, the validity of the results is
limited to the case in which the assumed models are the correct
representation of the true structure and dynamics of the cluster. The
second approach is more general, but we assume a constant $M/L$ ratio
and the resulting models could be non-physical (e.g. \cite[Binney \&
  Tremaine 1987]{BT87}).

In all cases we assumed spherical symmetry and velocity isotropy. This
seems to be reasonable for NGC 6388, which does not show significant
ellipticity (\cite[Harris 1996]){harris} and is thought to be quite
dynamically evolved (\cite[Ferraro et al. 2012]{dynclock}; see also
L11).  In both approaches we considered \cite[King (1966)]{king66} and
\cite[Wilson (1975)]{wilson75} distribution functions, which are
known to well reproduce the observed density profiles of GGCs (e.g.,
\cite[Mc Laughling \& van der Marel 2005]{mclvdm05}; {Miocchi et
  al. 2013}{miocchi13}).

As shown in Figures \ref{profs} and \ref{jeans}, our observed velocity
dispersion profile is consistent with both the absence of a central
IMBH, and the presence of a BH with a mass up to $\sim 2000 M_\odot$
(the latter possibility is mainly supported by the cusp in the
density profile). The absence of a central IMBH in NGC 6388 is
consistent with the results obtained from both X-ray and radio
observation, which put an upper limit of $\sim 600 M_\odot$ to a
possible compact dark mass in this cluster (\cite[Nucita et
  al. 2008]{nucita08}; \cite[Cseh et al. 2010]{cseh10}; \cite[Bozzo
  et al 2011]{bozzo11}).

\section{Summary}
We are using a multi-instrument approach (SINFONI+KMOS+FLAMES at the
ESO-VLT) to measure the radial velocities of hundred individual stars
distributed along the entire extension of $\sim 30$ Galactic GCs. The
shape of the innermost portion of the velocity dispersion profile (and
density distribution) will be used to constrain the presence of IMBHs
in these systems. The pilot project on NGC 6388 demonstrates that this
is the right route to properly determine GGC internal dynamics and
pursue the ``dynamical approach'' to the search for IMBHs. Our study
shows that no IMBH (or a $2000 M_\odot$ BH, at most) is required to
account for the observations in this cluster.

\vskip5truemm Most of the results discussed in this talk have been
obtained within the project Cosmic-Lab (PI: Ferraro, see
http:://www.cosmic-lab.eu), a 5-year project funded by the European
European Research Council under the 2010 Advanced Grant call (contract
ERC-2010-AdG-267675). I warmly thank the other team members involved
in this research: Francesco Ferraro, Elena Valenti, Alessio
Mucciarelli, Livia Origlia, Emilio Lapenna, Emanuele Dalessandro,
Michele Bellazzini, Paolo Miocchi, Davide Massari, Cristina Pallanca.

\end{document}